\documentclass[preprintnumbers, prd, twocolumn, showpacs,floatfix,preprintnumbers,
superscriptaddress, nofootinbib]{revtex4}
\usepackage{graphicx}
\usepackage{epsfig}
\usepackage{bm}
\usepackage{amssymb}
\usepackage{float}
\usepackage{amsmath}
\usepackage{subfigure}
\usepackage{dcolumn}
\usepackage{cancel}
\usepackage[colorlinks]{hyperref}
\usepackage[usenames,dvipsnames]{color}
\hypersetup{
     breaklinks=true,
    pdfstartview={FitH},    % fits the width of the page to the window
    colorlinks=true,       % false: boxed links; true: colored links
    linkcolor=blue,          % color of internal links
    citecolor=red,        % color of links to bibliography
    filecolor=magenta,      % color of file links
    urlcolor=blue,           % color of external links
    anchorcolor=green,      % Color for anchor text
    linktocpage=true
}

\def\doi{http://doi.org}

\begin{document}

\title{Reconstruction of interaction rate in
Holographic dark energy model with Hubble horizon as the infrared cut-off}

\author{Abdulla Al Mamon}
\email{abdulla.mamon@manipal.edu,\\
 abdulla.physics@gmail.com}
\affiliation{Manipal Centre for Natural Sciences, Manipal University, Manipal-576104, India}

\newcommand{\be}{\begin{equation}}
\newcommand{\ee}{\end{equation}}
\newcommand{\bea}{\begin{eqnarray}}
\newcommand{\eea}{\end{eqnarray}}
\newcommand{\bc}{\begin{center}}
\newcommand{\ec}{\end{center}}
%%%%%%%%%%%%%%%%%%%%%%%%%%%%%%%%%%%%%%%%%%%%%%%%%%%%%%%%%%%%%%%%%%%%%%%%%%
%%%%%%%%%%%%%%%%%%%%%%%%%%%%%%%%%%%%%%%%%%%%%%%%%%%%%%%%%%%%%%%%%%%%%%%%%%
\begin{abstract}
This work is the reconstruction of the interaction rate of holographic dark energy whose infrared cut-off scale is set by the Hubble length. We have reconstructed the interaction rate between dark matter and the holographic dark energy for a specific parameterization of the effective equation of state parameter. We have obtained observational constraints on the model parameters using the latest Type Ia Supernova (SNIa), Baryon Acoustic Oscillations (BAO) and Cosmic Microwave Background (CMB) radiation datasets. We have found that for the present model, the interaction rate increases with expansion and remains positive throughout the evolution. For a comprehensive analysis, we have also compared the reconstructed results of the interaction rate with other well-known holographic dark energy models. The nature of the deceleration parameter, the statefinder parameters and the dark energy equation of state parameter have also been studied for the present model. It has been found that the deceleration parameter favors the past decelerated and recent accelerated expansion phase of the universe. It has also been found that the dark energy equation of state parameter shows a phantom nature at the present epoch.
\end{abstract} 
%%%%%%%%%%%%%%%%%%%%%%%%%%%%%%%%%%%%%%%%%%%%%%%%%%%%%%%%%%%%%  
\maketitle 
\section{Introduction}
%%%%%%%%%%%%%%%%%%%%%%%%%%%%%%%%%%%%%%%%%%%%%%%%%%%%%%%%%%
Many cosmological observations such as Type Ia Supernova (SNIa)~\cite{SN,Riess:1998cb}, large scale structure (LSS)~\cite{LSS,Seljak:2004xh}, cosmic microwave background (CMB) radiation~\cite{Planck:2015xua, Ade:2015lrj, Ade:2014xna, Ade:2015tva, Array:2015xqh, Komatsu:2010fb, Hinshaw:2012aka} and baryon acoustic oscillations (BAO)~\cite{dvbc} have strongly confirmed that we live in an accelerating universe. All of these observations also strongly indicate that the universe was decelerating in the past and the observed cosmic acceleration is rather a recent phenomenon. In the literature, the most
accepted idea is that an exotic component of the matter sector with large negative pressure (i.e., long range anti-gravity properties), dubbed as “dark energy”, is responsible for this acceleration mechanism. Almost $68.3\%$ of the total energy budget of the universe at the present epoch is filled with dark energy (DE) while $26.8\%$ with dark matter (DM) and remaining $4.9\%$ being the usual baryonic matter and radiation \cite{planckr,planckr2,planckr3}. However, understanding the origin and nature of the dark sectors (DE and DM) is still a challenging problem in modern cosmology. For an excellent review on DE models, one can look into Refs. \cite{3,4,4a,5,6}. In the context of DE, the simplest and consistent with most of the observations is the cosmological constant model where the constant vacuum energy density serves as the DE candidate. However, the models based upon cosmological constant suffer from various inconsistencies, mainly the {\it fine tuning} and the cosmological {\it coincidence} problems \cite{sw, Steinhardt}. Thus, most of the recent research is aimed towards finding a suitable cosmologically viable model of dark energy. \\
%%%%%%%%%%%%%%%%%%%%%%%%%%%%%%%%%%%%%%%%%%%%%%%%%%%%%%%%%%%%%%%%%%%%%%%%%%%%%%%%
\par Most investigations in this direction are phenomenological and based on the assumption that the DE and DM evolve independently. However, one should realize that any interaction between the dark sectors will modify the evolution history of the universe. Presently, we neither know what the DM is made of nor do we understand the exact nature of DE. So, the possibility of DE to be in interaction with DM must be taken seriously. Recently, interacting DE models have gained immense interest and there are many works in which investigations are carried out considering an interaction between the different components of the universe (for review, see Refs. \cite{int1,int2,int3,int4,int5,int6,int7,int8,int9,int10,int11}).\\
%%%%%%%%%%%%%%%%%%%%%%%%%%%%%%%%%%%%%%%%%%%%%%%%%%%%%%%%%%%%%%%%%%%%%%%
\par This work mainly deals with the reconstruction of the interaction rate of {\it holographic dark energy} (HDE). In the context of DE, some earlier works on reconstruction can be found in \cite{rde1,rde2,rde3,rde4}, where various datasets have been used. The HDE model is based on the {\it holographic principle}, was proposed in Refs. \cite{hgm1,hgm2,hgm3,hgm4} by introducing the following energy density
\be
\rho = 3C^{2}M^2_{P}L^{-2} 
\ee
where $C$ is a constant to be determined by observational data. $M_{P}=(8\pi G)^{-\frac{1}{2}}$ and $L$ are the reduced Planck mass and the infrared cut-off scale, respectively. Recently, Campo et al. \cite{hgmls} have been studied holographic dark energy with different length scale cut-off such as the Hubble scale, the future event horizon and a scale proportional to the Ricci length. However, some previous works on HDE models have been comprehensively discussed in Refs. \cite{hder1,hder2,hder3,hder4,hder5}. In this paper, the Hubble horizon has been adopted as the infrared cut-off for the HDE meaning the cut-off length scale $L = H^{-1}$ , where $H$ is the Hubble parameter. Now it is important to note that the HDE models with Hubble horizon cut-off can generate late time cosmic acceleration along with the matter dominated decelerated expansion phase in the past only if there is some interaction between DM and DE \cite{sol1,sol0}. Reconstruction of interaction rate in HDE have earlier been discussed by several authors \cite{aasrdot,amq}, where the interaction rate has been reconstructed assuming a particular form of the DE equation of state (EoS) parameter \cite{aasrdot} or the deceleration parameter \cite{amq}. It is important to mention here that the interacting dark energy models with different forms of interaction term have been recently proposed by several authors in Refs. \cite{new1,new2}. In the current work, the interaction rate of HDE has been reconstructed from one specific parameterization of the {\it total or effective equation of state parameter} $\omega_{eff}(z)$. The basic properties of this chosen $\omega_{eff}(z)$ has been discussed in the next section. The main goal of the present work is to investigate the nature of interaction and the evolution of the interaction rate for this model assuming the HDE with Hubble horizon as the infrared cut-off. The observational constraints on the model parameters are obtained by using the latest SNIa, BAO and CMB datasets. Using the estimated values of model parameters, we have then reconstructed the interaction rate, the deceleration parameter and the dark energy equation of state parameter at the $1\sigma$ and $2\sigma$ confidence levels.\\
%%%%%%%%%%%%%%%%%%%%%%%%%%%%%%%%%%%%%%%%%%%%%%%%%%%%%%%%%%%%%%%%%%%%%%%%%%%
%%%%%%%%%%%%%%%%%%%%%%%%%%%%%%%%%%%%%%%%%%%%%%%%%%%%%%%%%%%%%%%%%%%%%%%%%%%%%%%%%%%%%%%%%
\par The outline of this paper is given as follows. In the next section, the reconstruction of the interaction rate for the present model has been discussed. In Section \ref{data}, we have described the latest observational datasets for our analysis, while in section \ref{result} we have used them to obtain the constraints
on the various quantities. The results of this analysis are also discussed in this section. Section \ref{conclusion} presents a brief summary of the work.
%%%%%%%%%%%%%%%%%%%%%%%%%%%%%%%%%%%%%%%%%%%%%%%%%%%%%%%%%%%%%%%%%%%%
%%%%%%%%%%%%%%%%%%%%%%%%%%%%%%%%%%%%%%%%%%%%%%%%%%%%%%%%%%%%%%%%
\section{Reconstruction of the interaction rate}\label{sec2}
%%%%%%%%%%%%%%%%%%%%%%%%%%%%%%%%%%%%%%%%%%%%%%%%%%%%%%%%%%%%%%%%%%%%%%%%%%%%%%%%%
We have considered the spatially flat FRW space-time 
\begin{equation}
ds^{2} = dt^{2} - a^{2}(t)[dr^{2} + r^{2}(d{\theta}^{2} + sin^{2}\theta d{\phi}^{2})]
\label{eq:2.2}
\end{equation}
where, $a(t)$ is the scale factor of the universe (which is considered to be $a = 1$ at the present epoch) and $t$ is the cosmic time. In a flat FRW background, the corresponding Einstein field equations can be obtained as,
\be\label{fe1}
H^{2}=\frac{8\pi G}{3}(\rho_{m}+\rho_{DE})
\ee
\be\label{fe2}
2{\dot{H}} + 3H^{2}=-8\pi G p_{DE}
\ee
where $H=\frac{\dot{a}}{a}$ is the Hubble parameter and an overhead dot implies differentiation with respect to time $t$. Here, $\rho_{m}$ represents the energy density of the dust (pressureless) matter while $\rho_{DE}$ and $p_{DE}$ represent the energy density and pressure of the dark energy. \\
\par As mentioned earlier, in the present work, our main aim is to study the interaction assuming a holographic dark energy with Hubble horizon ($H^{-1}$) as the infrared (IR) cut-off. Holographic dark energy models with the Hubble horizon as the IR cut-off require the interaction between the dark matter and dark energy to generate the late time acceleration. The dark energy density for a holographic model with the
Hubble horizon as the IR cut-off is given as \cite{sol0} 
\be 
\rho_{DE}=3C^{2}M^2_{P}H^{2}~~~~~~~~~({\rm Model}~1)
\ee
Here $C$ is a coupling parameter which is assumed to be a constant in the present work. Accordingly, the energy conservation equations become
\be
{\dot{\rho}}_{m}+3H\rho_{m}=Q 
\ee
\be 
{\dot{\rho}}_{DE}+3H(1+\omega_{DE})\rho_{DE}=-Q
\ee
where $\omega_{DE}=\frac{p_{DE}}{\rho_{DE}}$ is the equation of state parameter of dark energy and the $Q$ is the interaction term. If $Q=0$, then the matter evolve as, $\rho_{m}\propto a^{-3}$. Following Ref. \cite{sol1,sol0}, we have written the interaction as $Q=\rho_{DE}\Gamma$, where $\Gamma$ is an unknown function that measures the rate at which the energy exchange occurs between the dark sectors (DE and DM). The quantity of interest for analyzing the coincidence problem is the ratio $r=\frac{\rho_{m}}{\rho_{DE}}$ (known as coincidence parameter) and its time derivative can be written as \cite{aasrdot}
\be\label{eqrdot}
\dot{r}=(1+r)\Big[\Gamma + 3Hw_{DE}\frac{r}{1+r}\Big]
\ee
It should be noted that for a spatially flat universe, the coincidence parameter $r$ remains constant, irrespective of the value of $\frac{\Gamma}{H}$, for a holographic dark energy with Hubble horizon as the infrared cut-off. In this case, there will be no transition from decelerated to accelerated expansion \cite{sol1}. However, if the ratio $\frac{\Gamma}{H}$ is allowed to grow, then the transition may well occur, even though $r$ remains constant (for more details, see Refs. \cite{sol0,sol1}). For a constant value of $r$ (i.e., ${\dot{r}}=0$), the interaction rate can be obtained (using equation (\ref{eqrdot})) as
\be\label{eqintrate1}
\Gamma=-3Hr{\Big(\frac{\omega_{DE}}{1+r}\Big)}
\ee 
%%%%%%%%%%%%%%%%%%%%%%%%%%%%%%%%%%%%%%%%%%%%%%%%%%%%%%%%%%%%%%%
The effective or total EoS parameter can be expressed as,
\be\label{eqweff}
\omega_{eff}(z)=\frac{p_{eff}}{\rho_{eff}}=\frac{\omega_{DE}}{1+r} 
\ee
where, $\rho_{eff}=\rho_{m}+\rho_{DE}$ and $p_{eff}=p_{DE}$ are the effective energy density and effective pressure of the dark components (DM and DE) respectively.\\
Now the interaction rate can be written (in terms of $\omega_{eff}$) as
\be
\Gamma=-3Hr\omega_{eff} 
\ee
or
\be\label{eqintrate2}
\frac{\Gamma}{3H_0}=-{\left(\frac{H}{H_{0}}\right)}r\omega_{eff}
\ee
%%%%%%%%%%%%%%%%%%%%%%%%%%%%%%%%%%%%%%%%%%%%%%%%%%%%%%%%%%%%%%%
Clearly, we have freedom to choose one parameter as we have more unknown parameters with lesser numbers of equations (equations (\ref{fe1}) and (\ref{fe2})) to solve them. In the present work, we have considered a simple assumption regarding the functional form for the evolution of $\omega_{eff}(z)$ which is given by
%%%%%%%%%%%%%%%%%%%%%%%%%%%%%%%%%%%%%%%%%%%%%%%%%%%%%%%%%%%%%%%%%%%%%%%%%%%%%%%
\be\label{eans1}
\omega_{eff}(z)=-1 + \frac{A}{A+\frac{B}{(1+z)^n}}
\ee
where $A$, $B$ and $n$ are arbitrary constants. The system of equations is closed now. It deserves mention here that the effective EoS parameter $\omega_{eff}(z)$ at high redshift (i.e., $z\rightarrow \infty$) becomes effectively zero for any values of $A$ and $B$. On the other hand, it behaves like a dark energy at the present epoch, as $\omega_{eff}(z=0)=-\frac{B}{A+B}$ and thus depends on the values of $A$ and $B$. Thus the functional form of $\omega_{eff}(z)$ can easily accommodate both the phases of cosmic evolution (i.e.,  early matter dominated era and late-time dark energy dominated era). One advantage of this ansatz (as given in equation (\ref{eans1})) is that it is independent of any prior
assumption about the nature of dark energy. \\
%%%%%%%%%%%%%%%%%%%%%%%%%%%%%%%%%%%%%%%%%%%%%%%%%%%%%%%%%%%%%%%%%%%
\par Using equations (\ref{fe1}), (\ref{fe2}), (\ref{eqweff}) and (\ref{eans1}), the Hubble parameter for this model can be expressed as
\be
H(z)=H_{0}{\left[ \frac{B+A(1+z)^{n}}{A+B}\right]}^{\frac{3}{2n}}
\ee
where $H_{0}$ represents the present value of $H(z)$. The above equation can be re-written as
\be\label{eh}
h(z)={\left[ \alpha + (1-\alpha)(1+z)^{n}\right]}^{\frac{3}{2n}}
\ee
where, $h(z)=\frac{H(z)}{H_{0}}$ and for simplicity, we have defined $\alpha=\frac{B}{A+B}$, to be fixed by observations. It should be noted that the standard $\Lambda$CDM model corresponds to the case $n=3$, with present cold dark matter density parameter $\Omega_{m0}=(1-\alpha)$. Therefore, the model parameter $n$ is a good indicator of deviation of the present model from the $\Lambda$CDM model.\\ 
%%%%%%%%%%%%%%%%%%%%%%%%%%%%%%%%%%%%%%%%%
\par Now, the interaction rate of HDE can be reconstructed using the equations (\ref{eqintrate2}), (\ref{eans1}) and (\ref{eh}). It deserves mention here that for the reconstruction of the interaction rate, we need to fix the value of $r$. In this work, the value of $r$ is taken according to the recent measurement of the present DE density parameter $\Omega_{DE0}$ from Planck observation \cite{planckr} as for a spatially flat universe $r$ can be written as $r=\frac{(1-\Omega_{DE0})}{\Omega_{DE0}}$.\\
%%%%%%%%%%%%%%%%%%%%%%%%%%%%%%%%%%%%%%%%
%%%%%%%%%%%%%%%%%%%%%%%%%%%%%%%%%%%%%%%%%%%%%%%%%%%%%
\par The deceleration parameter $q$ is defined as
\be
q=-\frac{\ddot{a}}{aH^{2}}=-1+\frac{(1+z)}{H(z)}\frac{dH(z)}{dz} 
\ee
and for the present model, $q$ evolves as a function of $z$ as
\be
q(z)=-1 + \frac{3}{2}(1-\alpha)(1+z)^{n}{\left[ \alpha + (1-\alpha)(1+z)^{n}\right]}^{-1}
\ee
%%%%%%%%%%%%%%%%%%%%%%%%%%%%%%%%%%%%%%%%%%%%%%%%%%%%%%%%%%%%%%%%%%%%%%%%
\par Recently Sahni et al. \cite{rsp1,rsp2} have introduced a new geometrical diagnostic pair $\lbrace r,s \rbrace$ for dark energy, termed as ``statefinder parameters". These parameters can effectively differentiate between different models of dark energy and provide the best fit to existing observational data. These parameters are ( in terms of $H(z)$ and its derivatives ) \cite{rsp1,rsp2}
\be
r(z)=\frac{\dddot{a}}{aH^3}=1-2(1+z)\frac{H^{\prime}}{H}+ {\Big\lbrace \frac{H^{\prime \prime}}{H} + {\Big(\frac{H^{\prime}}{H}\Big)^2} \Big \rbrace}(1+z)^2
\ee
\be
s(z)=\frac{r(z)-1}{3(q-\frac{1}{2})}
\ee
In a flat $\Lambda$CDM model, the statefinder pair $\lbrace r,s \rbrace$ has a fixed point value and it is equal to $\lbrace 1, 0 \rbrace$. Thus, any deviation from $\lbrace 1, 0 \rbrace$ would favor a dynamical dark energy model. In this work, we have also studied the evolution of the parameter pair $\lbrace r,s \rbrace$ for the present model. However, one can look into Refs. \cite{new3,new4,new5}, where the authors have comprehensively discussed about the statefinder analysis for different dark energy models.\\
%%%%%%%%%%%%%%%%%%%%%%%%%%%%%%%%%%%%%%%%%%%%%%%%%%%%%%%
\par For a comprehensive analysis, in the present work, we have also compared our HDE model with the other versions of HDE (i.e., with other choices of IR cut-off length) by considering $i)$ Model $2:~L=R_{E}$, i.e., radius of the event horizon and $ii)$ Model $3:~L=({\dot{H}}+2H^2)^{-\frac{1}{2}}$, i.e., Ricci length scale (for review, see \cite{hgm4,hgmls} and references therein).\\
%%%%%%%%%%%%%%%%%%%%%%%%%%%%%%%%%%%%%%%%%%%%%
\par In the next section, we shall try to estimate the values of $\alpha$ and $n$ using SNIa, BAO and CMB datasets. With the help of those values of $\alpha$ and $n$, the ratio $\frac{\Gamma}{3H_{0}}$ can be reconstructed from the aforesaid datasets for each model (Models 1, 2 and 3).
%%%%%%%%%%%%%%%%%%%%%%%%%%%%%%%%%%%%%%%%%%%%%%%%%%%%%
\section{Data analysis methods}\label{data}
%%%%%%%%%%%%%%%%%%%%%%%%%%%%%%%%%%%%%%%%%%%%%%%%%
In this section, we have explained the data analysis method employed to constrain the theoretical model by using the recent observational datasets from Type Ia Supernova (SNIa), Baryon Acoustic Oscillations (BAO) and Cosmic Microwave Background (CMB) radiation data surveying. We have used the $\chi^{2}$ minimum test with these datasets and found the best fit values of
arbitrary parameters $\alpha$ and $n$ at the $1\sigma$ and $2\sigma$ confidence levels (as discussed in section \ref{result}). In the following subsections, the $\chi^{2}$ analysis used for those datasets is described.
%%%%%%%%%%%%%%%%%%%%%%%%%%%%%%%%%%%%%%%%%%%%%%%%%%%%%%%%%%%%%%%%%%%%%%%%%%%%%%%%%%%%%
\subsection{SNIa}
%%%%%%%%%%%%%%%%%%%%%%%%%%%%%%%%%%%%%%%%%%%%%%%%%%%%%%%%%%%%%%%%%%%%%%%%%%%%%%%%%%%%%%% 
The supernova distance modulus dataset of {\it joint light curve} analysis (JLA) has been utilized in the present work \cite{jla}. The binned dataset of JLA has been used along with the full covariance
matrix. The technical details of binning the data has been discussed comprehensively in reference \cite{jla}. The relevant $\chi^{2}$ is defined as \cite{jlam}
\be
\chi^{2}_{SNIa}=X(p)-\frac{Y^2(p)}{Z}-\frac{2{\rm ln}10}{5Z}Y(p)-Q^{\prime}
\ee
where
\bea
X(p)=\sum_{i,j}(\mu^{th}-\mu^{obs})_{i}({\cal C}ov)^{-1}_{ij}(\mu^{th}-\mu^{obs})_{j},\\
Y(p)=\sum_{i}(\mu^{th}-\mu^{obs})_{i}\sum_{j}({\cal C}ov)^{-1}_{ij},~~~~~~~~~~~~\\
Z=\sum_{i,j}({\cal C}ov)^{-1}_{ij}~~~~~~~~~~~~~~~~~~~~~~~~~~~~~~~~~~~~~~
\eea
Here, $``{\cal C}ov"$ is the covariance matrix of the binned data sample, $Q^{\prime}$ is a constant that does not depend on the model parameter $p$ and hence has been ignored. Also, $\mu^{obs}$ represents the observed distance modulus, while $\mu^{th}$ is for the theoretical one, which is defined as
\be
\mu^{th}=5{\rm log}_{10}{\Big[\frac{d_{L}(z)}{1{\rm Mpc}}\Big]}+ 25
\ee
where, 
\be
d_{L}(z)=\frac{(1+z)}{H_{0}}\int^{z}_{0}\frac{dz^{\prime}}{h(z^{\prime})} 
\ee
%%%%%%%%%%%%%%%%%%%%%%%%%%%%%%%%%%%%%%%%%%%%%%%%%%%%%%%
%%%%%%%%%%%%%%%%%%%%%%%%%%%%%%%%%%%%%%%%%%%%%%%%%%%
\subsection{BAO/CMB}
%%%%%%%%%%%%%%%%%%%%%%%%%%%%%%%%%%%%%%%%%%%%%%%%%%%%
In this section, we have briefly discussed the combined BAO/CMB dataset. We start by defining the comoving sound horizon at the photon-decoupling epoch
\begin{equation}
r_s(z_*)= \frac{c}{\sqrt{3}} \int_0^{\frac{1}{(1+z_*)}}\frac{da}{a^2H(a)\sqrt{1+\Big(\frac{3\Omega_{b0}}{4 \Omega_{\gamma 0}}\Big)a} } 
\end{equation}
where $\Omega_{b0}$ and $\Omega_{\gamma 0}$ are, respectively, the present value of the  baryon and photon density parameter and $z_*$ is the redshift of photon decoupling. To obtain the BAO/CMB constraints another relevant quantity is the redshift of the drag epoch ($z_d$), when the photon pressure is no longer able to avoid gravitational instability of the baryons. The Planck 2015 \cite{Planck:2015xua} values for these two redshifts are  $z_*=1090.00\pm 0.29$ and $z_d=1059.62\pm 0.31$. We have also used the ``acoustic scale''
$l_A=\pi\frac{d_A(z_*)}{r_s(z_*)}$, and  the ``dilation scale'',  
$D_V(z)=\left[ \frac{d_A^2(z) cz}{H(z)} \right]^{1/3}$, introduced in Ref. \cite{dvbc}. Here,  $d_A(z_*)= c \int_{0}^{z_*}\frac{dz^{\prime}}{H(z^{\prime})}$ is the comoving angular-diameter distance. In this work, for BAO dataset, we have used the results from Padmanabhan et al. \cite{padmanabhan12}, which reported a measurement of $\frac{r_s}{D_V}$ at $z=0.35$, ($\frac{r_s(z_d)}{D_V(0.35)}=0.1126\pm0.0022$), Anderson et al. \cite{anderson13}, ($\frac{r_s(z_d)}{D_V(0.57)}=0.0732\pm0.0012$), Beutler et al. \cite{beutler11}, ($\frac{r_s(z_d)}{D_V(0.106)}=0.336\pm0.015$) and Blake et al. \cite{blake11}, which obtained results at $z=0.44$, $z=0.6$ and $z=0.73$ ($\frac{r_s(z_d)}{D_V(0.44)}=0.0916\pm0.0071$, $\frac{r_s(z_d)}{D_V(0.6)}=0.0726\pm0.0034$ and $\frac{r_s(z_d)}{D_V(0.73)}=0.0592\pm0.0032$). In addition to this, we have combined theses results with the CMB measurement derived from the Planck 2015 \cite{Planck:2015xua} observational values $l_A = \frac{\pi}{\theta_*}= 301.77\pm0.09$, $r_s(z_d)=147.41\pm 0.30$ and $r_s(z_*)=144.71\pm 0.31$, for the combined analysis TT, TE, EE+lowP+lensing. However, the details
of the methodology for obtaining the constraints on model
parameters using the the BAO/CMB dataset can be found in Ref. \cite{bcp15}.\\
%%%%%%%%%%%%%%%%%%%%%%%%%%%%%%%%%%%%%%%%%%%%%%%%%%%%%%%%%%%%%%%%%%%%%%%
The $\chi^2$ function for this dataset is defined as
\begin{equation}
\chi^2_{BAO/CMB}={\bf X^{T}C^{-1}X},
\label{chi2baocmb}
\end{equation}
where
\begin{equation}
{\bf X}=\left(
          \begin{array}{cccc}
          \displaystyle\frac{d_A(z_*)}{D_V(0.106)} -30.84 \\
            \displaystyle\frac{d_A(z_*)}{D_V(0.35)} -10.33 \\
            \displaystyle\frac{d_A(z_*)}{D_V(0.57)} -6.72 \\
             \displaystyle\frac{d_A(z_*)}{D_V(0.44)} -8.41 \\
             \displaystyle\frac{d_A(z_*)}{D_V(0.6)} -6.66 \\
              \displaystyle\frac{d_A(z_*)}{D_V(0.73)} -5.43 \\
          \end{array}
        \right)
\end{equation}
and the inverse covariance matrix ($C^{-1}$) can be found in Ref. \cite{bcp15}. \\
%%%%%%%%%%%%%%%%%%%%%%%%%%%%%%%%%%%%%%%%%%%%%%%%%%%%%
\par Now, one can use the maximum likelihood method and take the total likelihood function as
\be 
{\cal L}={\rm e}^{-\frac{\chi^2}{2}}
\ee
where, $\chi^{2}=\chi^{2}_{SNIa}+\chi^{2}_{BAO/CMB}$.
For the SNIa+BAO/CMB dataset, one can obtain the best-fit values of parameters by minimizing $\chi^2$. The best-fit parameter values $p^{*}$ are those that maximize the likelihood function. Consequently, the $1\sigma$ and $2\sigma$ confidence level contours correspond to the sets of parameters (centered on $p^{*}$) bounded by $\chi^2(p^{*}) + 2.3$ and $\chi^2(p^{*}) + 6.17$ respectively. In this work, we have minimized the $\chi^{2}$ function (say, $\chi^2_{min}$) with respect to the model parameters $\lbrace \alpha ,n\rbrace$ to estimate their best fit values.
%%%%%%%%%%%%%%%%%%%%%%%%%%%%%%%%%%%%%%%%%%%%%%%%%%%%%%%%%
\section{Results}\label{result} 
%%%%%%%%%%%%%%%%%%%%%%%%%%%%%%%%%%%%%%%%%%%%%%%%%%%%%%%%%%%%%%%%%%%%%%%%
In this section, we have obtained the constraints on the model parameters ($\alpha$ and $n$) for the combined dataset (SNIa+BAO/CMB) and have discussed the results obtained from the statistical analysis. The corresponding $1\sigma$ and $2\sigma$ confidence level contours in $\alpha - n$ plane is shown in figure {\ref{figc} for this dataset. The best-fit values for the model parameters are obtained as, $\alpha=0.74^{+0.01}_{-0.02}~(1\sigma)$ and $n=3.27^{+0.20}_{-0.19}~(1\sigma)$ with $\chi^2_{min}=32.78$.
%%%%%%%%%%%%%%%%%%%%%%%%%%%%%%%%%%%%%%%%%%%%%%%%
\begin{figure}[ht]
\begin{center}
\includegraphics[width=0.35\textwidth,height=0.2\textheight]{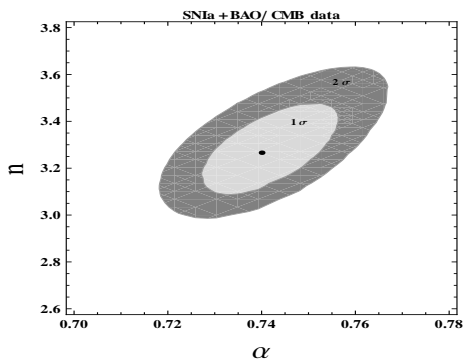}
\caption{Plot of $1\sigma$ (light gray) and $2\sigma$ (gray) confidence contours on $\alpha-n$ parameter space. In this plot, black dot represents the best-fit values of $\alpha$ and $n$ arising from the analysis of SNIa+BAO/CMB dataset. }
\label{figc}
\end{center}
\end{figure}
%%%%%%%%%%%%%%%%%%%%%%%%%%%%%%%%%%%%%%%%%%%%%%%%%%%%%%%%%%
Also, the marginalized likelihoods for the present model is shown in figure \ref{figl}. It is clear from figure \ref{figl} that the likelihood functions are well fitted
to a Gaussian distribution function for the combined dataset.\\
%%%%%%%%%%%%%%%%%%%%%%%%%%%%%%%%%%%%%%%%%%%%%%%%%%%%%%%%%%%%%
\begin{figure}[ht]
\begin{center}
\includegraphics[width=0.38\textwidth,height=0.21\textheight]{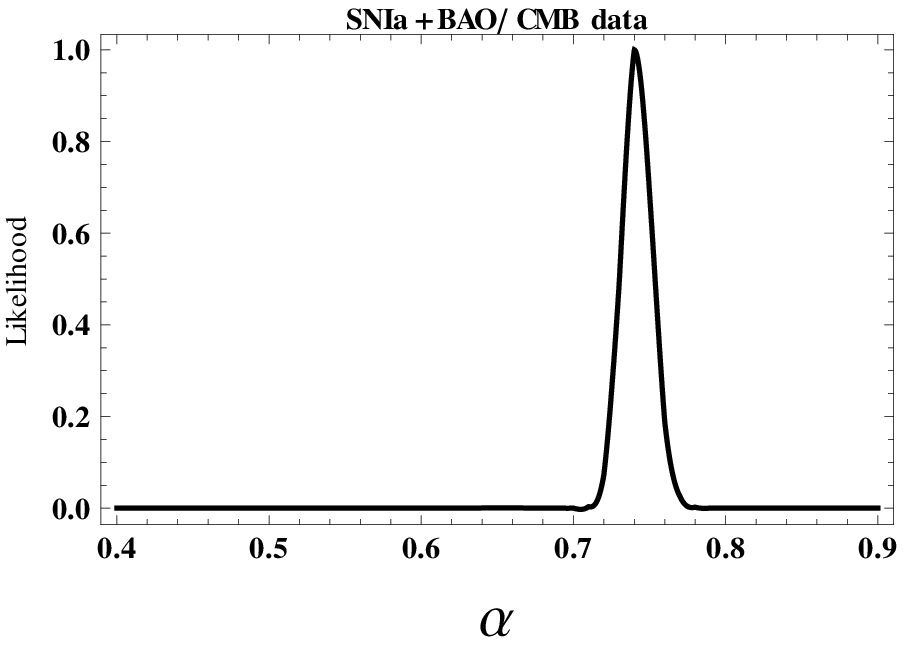}
\hspace{5mm}
\includegraphics[width=0.38\textwidth,height=0.21\textheight]{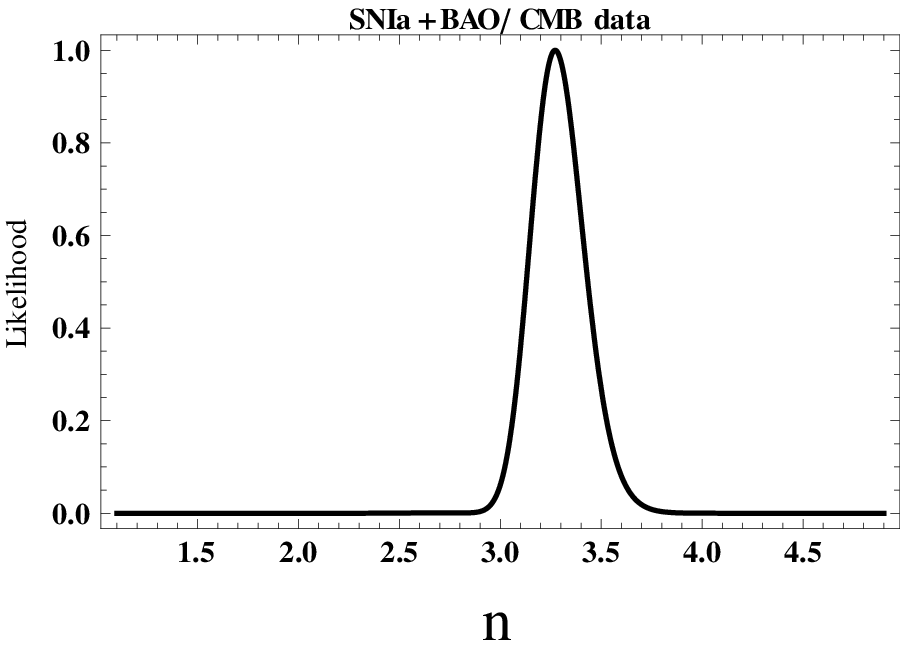}
\caption{The marginalized likelihood functions of the present model are shown for SNIa+BAO/CMB dataset.}
\label{figl}
\end{center}
\end{figure}
%%%%%%%%%%%%%%%%%%%%%%%%%%%%%%%%%%%%%%%%%%%%%%%%%%%%%%%
\par Figure \ref{figintr} shows the nature of constraint on the interaction rate $\frac{\Gamma}{3H_{0}}$ for Models 1, 2 and 3, obtained in the analysis of SNIa+BAO/CMB dataset. It is evident from figure \ref{figintr} that the interaction rate increases at recent time, but it was significantly low at earlier time. For Model 1, the present value of $\frac{\Gamma}{3H_{0}}$ is found to be $0.339^{+0.006}_{-0.006}$ at the $2\sigma$ confidence level and also the interaction rate  remains positive throughout the evolution. Consequently, the interaction term $Q$ is also positive through the evolution (as $Q=\rho_{DE}\Gamma$) and thus the energy transfers from DE to DM which is well consistent with the second law of thermodynamics and the Le Chatelier-Braun principle \cite{2lt}. It should be noted that similar results have also been obtained by Sen and Pavon \cite{aasrdot}, where the interaction rate of HDE has been reconstructed from a parametrization of DE equation of state parameter. However, in the present work, we have obtained more tighter constraints as compared to the previous findings. On the other hand, we have also found from figure \ref{figintr} that the Model 2 and Model 3 are not always consistent with the second law of thermodynamics and the Le Chatelier-Braun principle (as $\frac{\Gamma}{3H_{0}}$ is not positive throughout the evolution), which is in contrast to the result obtained in Model 1.\\
%%%%%%%%%%%%%%%%%%%%%%%%%%%%%%%%%%%%%%%%%%%%%%%%
\begin{figure}[ht]
\begin{center}
\includegraphics[width=0.4\textwidth,height=0.2\textheight]{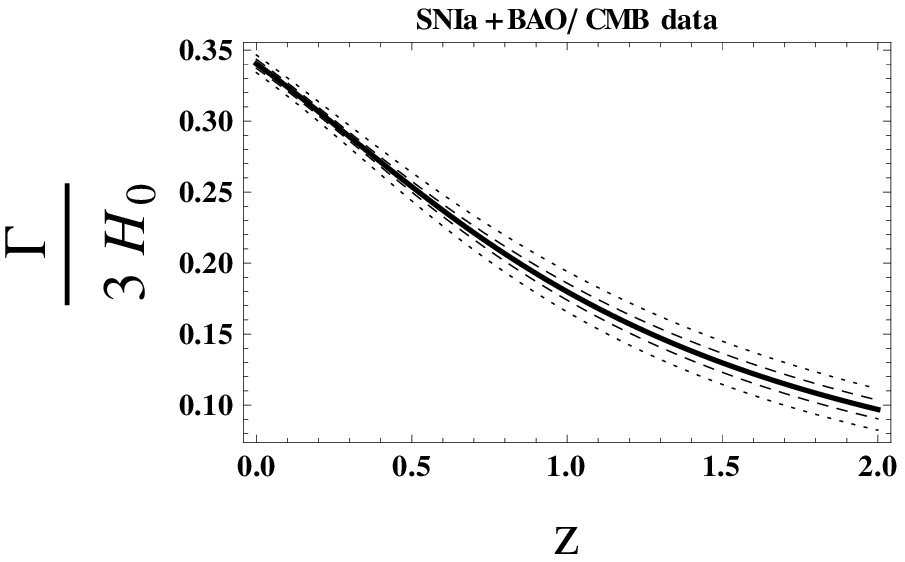}
\hspace{5mm}
\includegraphics[width=0.4\textwidth,height=0.2\textheight]{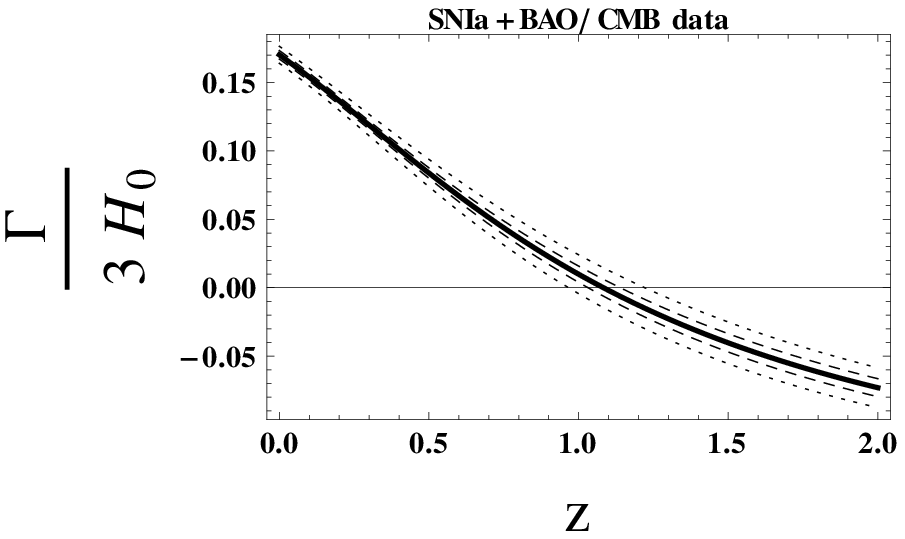}\\
\includegraphics[width=0.4\textwidth,height=0.2\textheight]{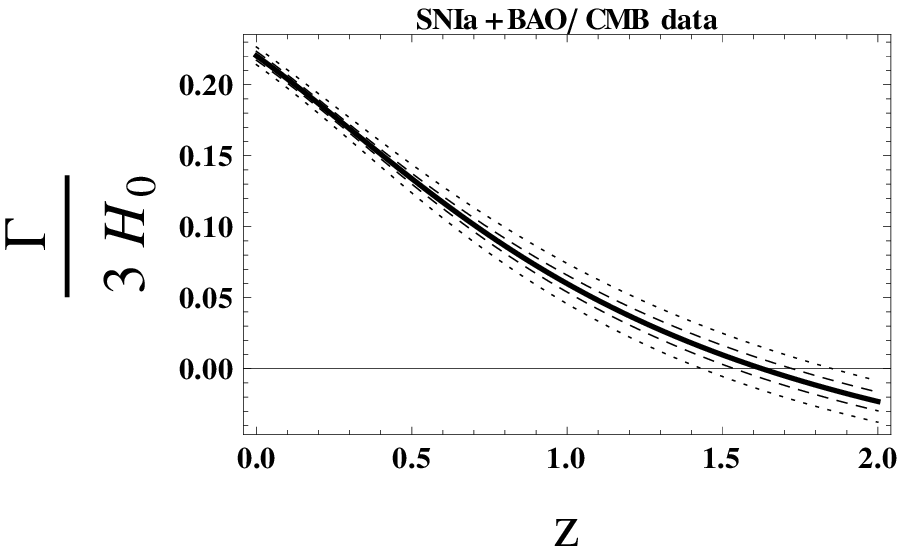}
\caption{The plot of $\frac{\Gamma}{3H_{0}}$ vs. $z$ is shown for Model 1 (upper panel), Model 2 (middle panel) and Model 3 (lower panel) by considering the SNIa+BAO/CMB dataset. The dashed and dotted curves represent the $1\sigma$ and $2\sigma$ confidence contours respectively, while the central thick line represents the best-fit curve.}
\label{figintr}
\end{center}
\end{figure}
%%%%%%%%%%%%%%%%%%%%%%%%%%%%%%%%%%%%%%%%%%%%%%%%%%%%
%%%%%%%%%%%%%%%%%%%%%%%%%%%%%%%%%%%%%%%%%%%%%%%%%%%%%%%%%%%%%
\begin{figure}[ht]
\begin{center}
\includegraphics[width=0.38\textwidth,height=0.21\textheight]{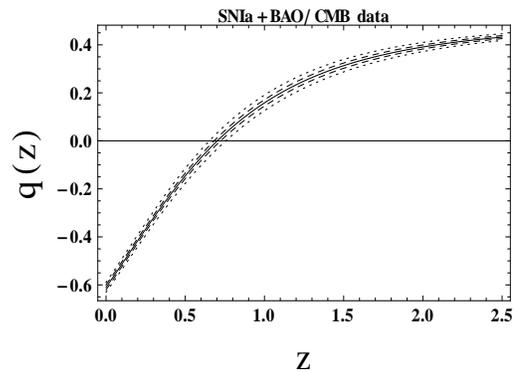}
\caption{The evolution of $q(z)$ is shown within $1\sigma$ (dashed curve) and $2\sigma$ (dotted curve) confidence regions around the best fit curve (central thin curve) for the present model. This is for SNIa+BAO/CMB dataset. }
\label{figqz}
\end{center}
\end{figure}
%%%%%%%%%%%%%%%%%%%%%%%%%%%%%%%%%%%%%%%%%%%%%%%%%%%%%%%
%%%%%%%%%%%%%%%%%%%%%%%%%%%%%%%%%%%%%%%%%%%%%%%%%%%%%%%%%%%%%
\begin{figure}[ht]
\begin{center}
\includegraphics[width=0.38\textwidth,height=0.21\textheight]{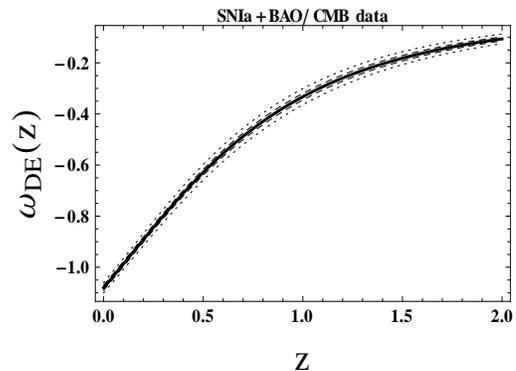}
\caption{The evolution of $\omega_{DE}(z)$ is shown within $1\sigma$ (dashed curve) and $2\sigma$ (dotted curve) confidence regions around the best fit curve (central thin curve) for the present model.}
\label{figwde}
\end{center}
\end{figure}
%%%%%%%%%%%%%%%%%%%%%%%%%%%%%%%%%%%%%%%%%%%%%%%%%%%%%%%
%%%%%%%%%%%%%%%%%%%%%%%%%%%%%%%%%%%%%%%%%%%%%%%%%%%%%%%%%%%%%
\begin{figure}[ht]
\begin{center}
\includegraphics[width=0.38\textwidth,height=0.21\textheight]{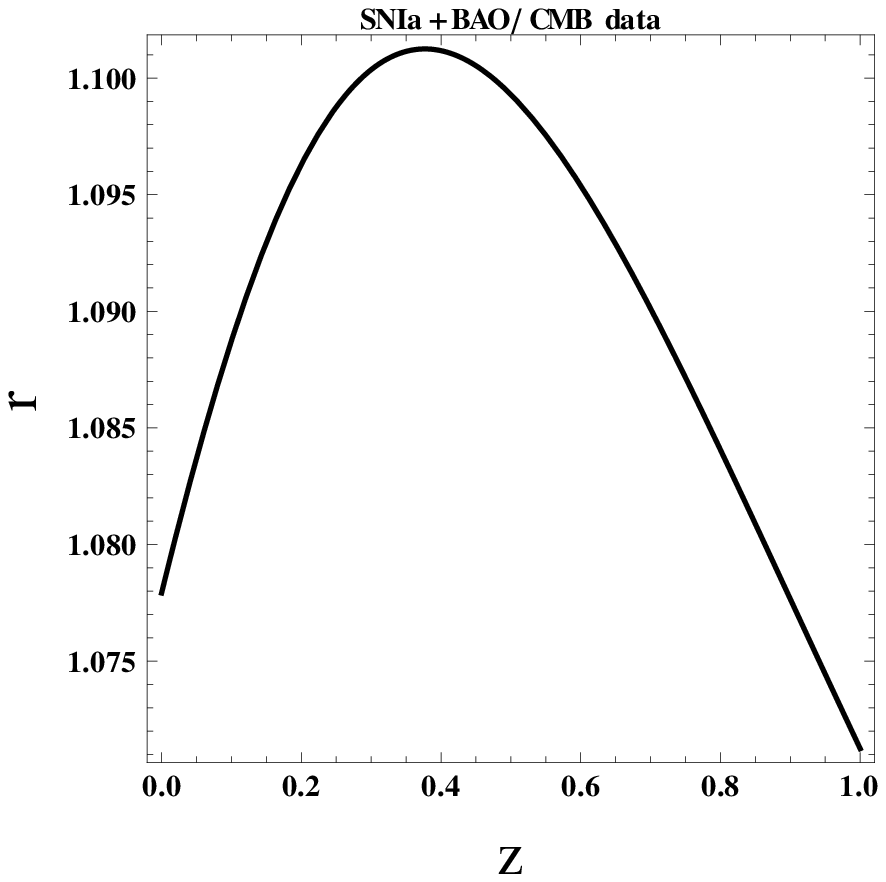}\hspace{7mm}
\includegraphics[width=0.38\textwidth,height=0.21\textheight]{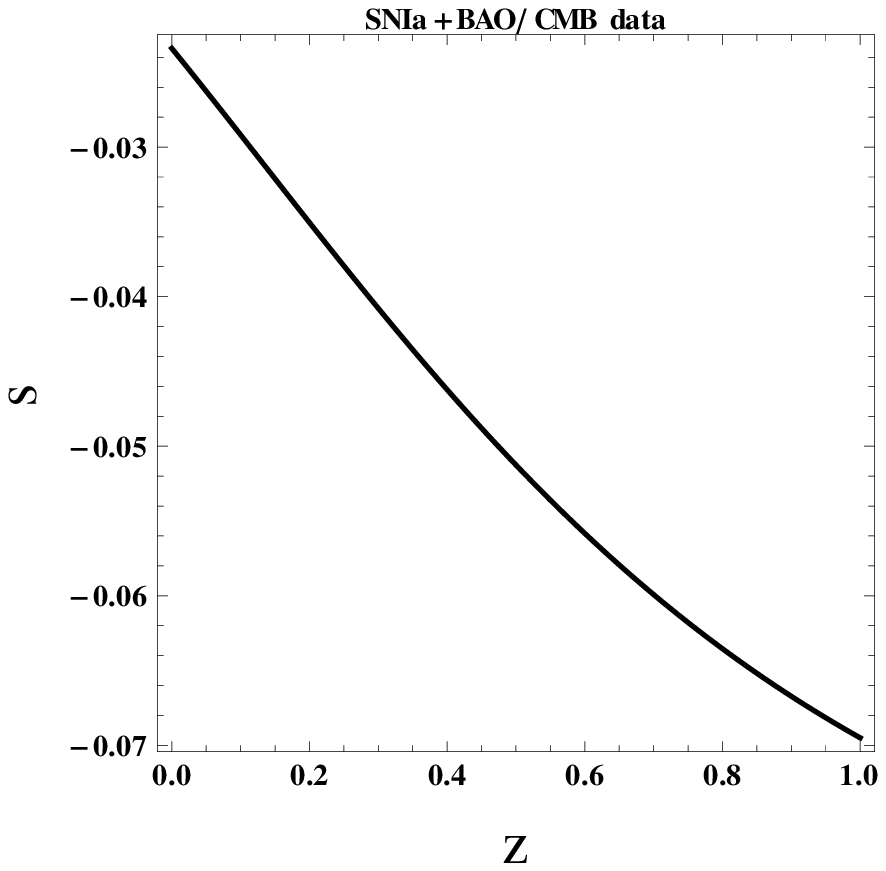}
\caption{The upper panel shows the evolution of the statefinder $r$ for the best fit model, where as the lower panel shows the evolution of the statefinder $s$.}
\label{figrs}
\end{center}
\end{figure}
%%%%%%%%%%%%%%%%%%%%%%%%%%%%%%%%%%%%%%%%%%%%%%%%%%%%%%%
%%%%%%%%%%%%%%%%%%%%%%%%%%%%%%%%%%%%%%%%%%%%%%%%%%%%%%%%%%
\par For a completeness, we have also reconstructed the evolutions of $q(z)$ and $\omega_{DE}(z)$ for the present model. In figure \ref{figqz}, the evolution of $q(z)$ is shown within $1\sigma$ and $2\sigma$ confidence regions around the best fit curve for the SNIa+BAO/CMB dataset. It is evident from figure \ref{figqz} that $q(z)$ shows a smooth transition from a decelerated ($q > 0$) to an accelerated ($q < 0$) phase of expansion of the universe at the transition redshift $z_{t}=0.7^{+0.02}_{-0.02}$ (within $1\sigma$ errors) and $z_{t}=0.7^{+0.05}_{-0.05}$ (within $2\sigma$ errors) for the best-fit model. This results are found to be consistent with the results obtained independently by several authors (see \cite{zt1,zt2,zt3} and references there in), which states that the universe at redshift in between $z_{t}\sim 0.5-1$  underwent a phase transition from decelerating to accelerating expansion. Furthermore, we have also reconstructed the EoS parameter $\omega_{DE}(z)$ for the dark energy in figure \ref{figwde}. We have found that for the best-fit model, the present value of $\omega_{DE}(z)$ comes out to be $-1.07^{+0.01}_{-0.01}$ (with $1\sigma$ errors) and $-1.07^{+0.02}_{-0.02}$ (with $2\sigma$ errors) and thus shows a phantom nature ($\omega_{DE}(z=0)<-1$) at the present epoch. This result is also in good agreement with the recent observational constraints on $\omega_{DE}$ ($-1.1\le \omega_{DE}\le -0.9$) obtained by Wood-Vasey et al. \cite{wv} and Davis et al. \cite{dav}. It should be noted that the best-fit value of $\omega_{DE}$ do not deviate very much from that of the Planck 2015 analysis, which puts the limit on the parameter as, $\omega_{DE}=-1.006\pm 0.045$ \cite{Planck:2015xua}. It has also been found from figure \ref{figwde} that $\omega_{DE}(z)$ tends to zero at high redshift and it behaves like dust matter in the past. This result allows a matter dominated epoch in the recent past. As mentioned earlier that the parameter $n$ is a good indicator of deviation of the present model from the standard $\Lambda$CDM model as for $n=3$ the model mimics the $\Lambda$CDM. It is always good to have a $\Lambda$CDM value ($\omega_{\Lambda}=-1$) for a model to be consistent with observations, but as the true nature of dark energy is still unknown, this slight deviations from the $\Lambda$CDM value also need attention. Furthermore, the best fit evolution of the statefinder pair $\lbrace r,s\rbrace$ as a function of $z$ is shown in figure \ref{figrs}, which indicates the departure of $\lbrace r,s\rbrace$ from the standard $\Lambda$CDM model (for which $r=1$ and $s=0$). However, it is seen that our model $( r=1.078,s=-0.02)$ does not deviate very far from the $\Lambda$CDM model at the present epoch.   
%%%%%%%%%%%%%%%%%%%%%%%%%%%%%%%%%%%%%%%%%%%%%%%%%%%%%%%%%%%%%%
\section{Conclusions}\label{conclusion}
%%%%%%%%%%%%%%%%%%%%%%%%%%%%%%%%%%%%%%%%%%%%%%%%%%%%%%%
As mentioned earlier, this work is the reconstruction of the interaction rate of  holographic dark energy (HDE) whose infrared cut-off scale is set by the Hubble length. The HDE model, discussed in this paper, is based on the parameterization of the total equation of state parameter. We have reconstructed the interaction rate $\Gamma$ (in terms of $3H_{0}$) by combining the observations from SNIa, BAO and CMB. As discussed earlier, we have also compared our model with two other HDE models to draw a direct comparison between them. It is clear from figure \ref{figintr} that the interaction rate increases at recent time from a small value in the long past. It has also been found that for the Model 1, the interaction rate (and hence $Q$) remains positive throughout the evolution. This suggests that the energy transfers from DE to DM, which is well consistent with the second law of thermodynamics and the Le Chatelier-Braun principle \cite{2lt}.\\
\par We have shown that the deceleration parameter undergoes a smooth transition from a positive value to some negative value which indicates that the universe was undergoing an early deceleration followed by late time acceleration. This is essential for the structure formation of the universe. In addition, the best-fit value of the transition redshift ($z_{t}=0.7^{+0.05}_{-0.05}$, at $2\sigma$ level) obtained in this work is found to be consistent with the results obtained independently by several authors \cite{zt1,zt2,zt3}. Furthermore, it has also been found that for the best-fit model, the evolution of $\omega_{DE}$ shows a little phantom nature at the present epoch, which is well consistent with the Planck 2015 data \cite{Planck:2015xua}. Therefore, this particular HDE model (with $Q>0$ and $\omega_{DE}(z=0)\approx -1$) can be considered as an alternative for the standard $\Lambda$CDM model.
%%%%%%%%%%%%%%%%%%%%%%%%%%%%%%%%%%%%%%%%%%%%%%%%%%%%%%%%%%%%%%%%%%%%%
%%%%%%%%%%%%%%%%%%%%%%%%%%%%%%%%%%%%%%%%%%%%%%%%%%%%%%%%%%
%%%%%%%%%%%%%%%%%%%%%%%%%%%%%%%%%%%%%%%%%%%%%%%%
\section{Acknowledgments}
The author is thankful to the anonymous referee whose valuable comments have improved the quality of this paper.
%%%%%%%%%%%%%%%%%%%%%%%%%%%%%%%%%%%%%%%%%%%%%%%%%%%%%%%%%%%%%%%%%%%%%
%%%%%%%%%%%%%%%%%%%%%%%%%%%%%%%%%%%%%%%%%%%%%%%%%%%%%%%%%%%%%%%

\end{document}